# Towards Deep Learning Surrogate for the Forward Problem in Electrocardiology: A Scalable Alternative to Physics-Based Models

Shaheim Ogbomo-Harmitt[1]*, Cesare Magnetti[2]*, Chiara Spota[1], Jakub Grzelak[1], Oleg Aslanidi[1]

[1] School of Biomedical Engineering and Imaging Sciences, King's College London, London, UK
[2]PhysicsX, London, UK

## Abstract

*The forward problem in electrocardiology, computing body surface potentials from cardiac electrical activity, is traditionally solved using physics-based models such as the bidomain or monodomain equations. While accurate, these approaches are computationally expensive, limiting their use in real-time and large-scale clinical applications. We propose a proof-of-concept deep learning (DL) framework as an efficient surrogate for forward solvers. The model adopts a time-dependent, attention-based sequence-to-sequence architecture to predict electrocardiogram (ECG) signals from cardiac voltage propagation maps. A hybrid loss combining Huber loss with a spectral entropy term was introduced to preserve both temporal and frequency-domain fidelity. Using 2D tissue simulations incorporating healthy, fibrotic, and gap junction–remodelled conditions, the model achieved high accuracy (mean $R^2$ = 0.99 ± 0.01). Ablation studies confirmed the contributions of convolutional encoders, time-aware attention, and spectral entropy loss. These findings highlight DL as a scalable, cost-effective alternative to physics-based solvers, with potential for clinical and digital twin applications.*

## 1. Introduction

The forward problem in electrocardiology, evaluating body surface potentials from cardiac electrical activity, is traditionally addressed using physics-based models, such as the Bidomain model, typically solved via the finite element method[1, 2]. While these approaches are known for their relative effectiveness, they are computationally intensive. Limiting their practicality in time-sensitive or large-scale clinical applications. Recent research has leveraged these models to calibrate patient-specific simulations using electrocardiogram (ECG) data, thereby enhancing the personalisation of cardiac digital twins [3]. However, widespread clinical adoption necessitates more scalable and computationally efficient alternatives. Beyond personalisation, solving the forward problem has been applied in guiding therapeutic interventions. Krummen et al. have generated a library of cardiac tissue electrical simulations with corresponding ECGs to localise arrhythmic sites of origin in both atrial and ventricular cases based on 12-lead ECG data [4].

This demonstrates the dual utility of solving the forward problem: (i) facilitating routine clinical deployment, and (ii) enabling the development of large-scale simulation libraries to support targeted treatments. To address these needs, deep learning (DL) could be a promising approach, offering significant reductions in computational burden. In this study, we present a proof-of-concept DL framework for the forward problem in electrocardiology. Specifically, we introduce a time-dependent, attention-based sequence-to-sequence model designed to predict ECG signals from sequences of 2D cardiac voltage propagation maps. The proposed approach is intended to provide a cost-effective and scalable surrogate to conventional forward problem solvers.

## 2. Methods

### 2.1. 2D simulation

A dataset of 300 2D tissue models (200 x 200 mm) was generated and used to simulate cardiac electrical activity by solving the monodomain equation with an atrial variant of the Fenton-Karma cell model [5]. Cardiomyopathy was modelled as a variable distribution of non-conductive fibrotic tissue, with different diffusion coefficients ($\boldsymbol{D}$) representing tissue conductivity. Four distinct tissue types were simulated: 60 healthy cases ($\boldsymbol{D}$ = 0.10–0.09 mm²s$^{-1}$) with small non-conductive tissue patches; 60 gap junctional remodelling cases ($\boldsymbol{D}$ = 0.09–0.01 mm²s$^{-1}$); 120 fibrotic remodelling cases with large non-conductive patches; and 60 combined remodelling cases. To compute extracellular potential (ECG) signals for each simulation, the following equation was used:

$$\phi_e = \iiint \frac{\nabla . \boldsymbol{D} \nabla V_m}{|\boldsymbol{r} - \boldsymbol{r}_e|} d\boldsymbol{r} \qquad (1)$$

Where $\phi_e$ is the extracellular potential at the observation point $\boldsymbol{r}_e$, $V_m$ is the transmembrane potential, $\boldsymbol{D}$ is the diffusion coefficient and $\boldsymbol{r}$ is the spatial coordinates of cardiac tissue domain [6]. The electrode was positioned at spatial coordinates (50, 50, 40) mm, defined relative to the origin located at the top-left corner of the square tissue domain.

## 2.2. Model

The proposed deep-learning framework employs a sequence-to-sequence architecture: comprising an encoder, an attention mechanism and a decoder to translate a series of voltage propagation maps into an output signal (Figure 1). Each input map $x_t \in \mathbb{R}^{C\times H\times W}$ is processed by the encoder, a two-block convolutional network with batch normalisation, ReLU activations, pooling and dropout, followed by adaptive pooling, flattening and a fully connected layer to yield a latent vector $l_t \in \mathbb{R}^{T\times hidden\ size}$. These latent representations $L = \{l_1, l_2, \dots, l_t\}$ are then weighted by the attention module and decoded into the final output.

The attention mechanism uses Bahdanau additive attention enhanced with sinusoidal time embeddings to produce a context vector $c_t$. At each decoding step $t$, attention weights $a_{t,i}$ are computed from the current decoder state $h_t$, each latent vector $l_i$, and its time embedding $e_i^a$ (Equation 2) [7]. By integrating explicit temporal information, this time-aware attention lets the model prioritise latent vectors according to both their voltage content and their time steps, thereby strengthening its capture of sequential dependencies in the forward electrocardiology task.

$$a_{t,i} = V_a^{\mathrm{T}} \tanh\left(W_a h_t + U_a(l_i \parallel e_i^a)\right) \tag{2}$$

$W_a$, $U_a$ and $V_a$ represent learned attention parameters, and $\parallel$ indicates concatenation. Attention weights, $\alpha$, are then evaluated by applying the SoftMax function to the attention scores $a$, were $\alpha_{t,i} \in [0,1]$. Finally, the context vector, $c_t$, is computed as the weighted sum of the $L$ with weights $\alpha$ (Equation 3).

$$c_t = \sum_i \alpha_{t,i} l_i\,, \quad where\ c_t \in \mathbb{R}^{hidden\ size} \tag{3}$$

At each decoding step, the long short-term memory (LSTM) decoder receives the previous hidden and cell states, along with a concatenated input consisting of the attention-derived context ($c_t$), the encoder's latent vector at the current time step ($l_t$), and a sinusoidal time embedding ($e_t^d$) indicating the output index. The LSTM output is then combined with the previous ECG prediction and passed through two linear layers to produce the next ECG value. This structure enables the model to generate temporally consistent predictions while explicitly leveraging both input and output timing information.

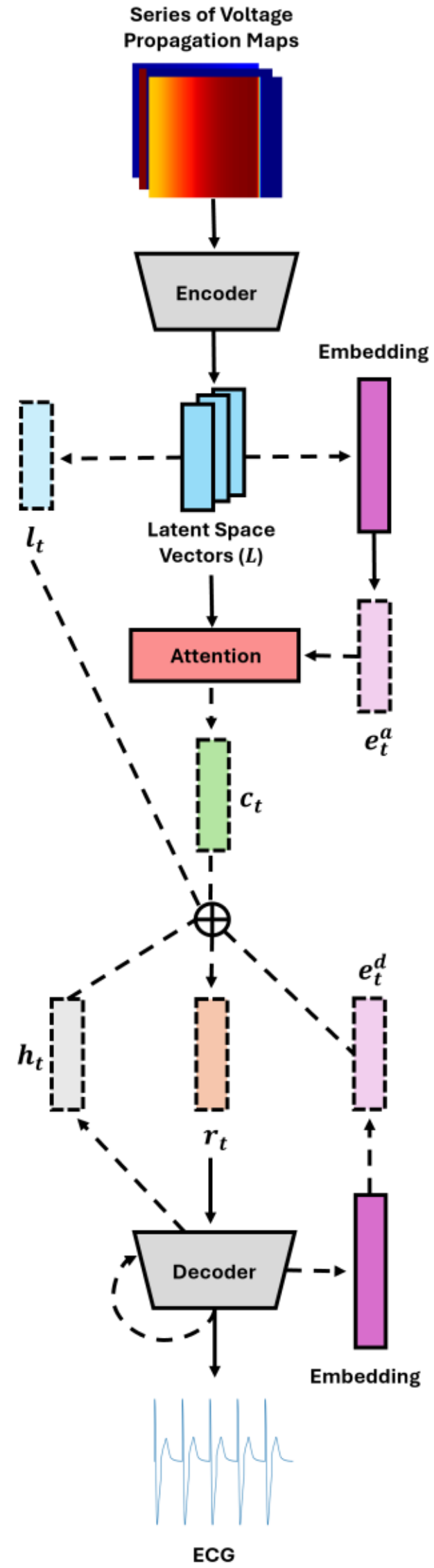

**Figure 1.** Schematic of DL model architecture from series of 2D wave propagation maps to ECG prediction. Solid lines show static data flow, and dashed lines show recurrent data flow.

## 2.3. Model Training and Evaluation

To train the model, we introduce a novel loss function that jointly minimises differences in both the time and frequency domains. Frequency-domain differences are captured using a spectral entropy loss, which encourages preservation of the signal's spectral complexity. Given a ground truth ECG signal $y_i \in \mathbb{R}^T$ and a predicted signal $\widehat{y}_i \in \mathbb{R}^T$, where $i$ indexes samples in a batch of size $B$, we compute the power spectral density (PSD) for each signal $s \in \{\hat{y}_i, y_i\}$, denoted $P_s(f) \in \mathbb{R}^T$. The PSD is normalised into a probability distribution by dividing each spectral component by the total power across all frequencies, such that $\widetilde{P}_s(f) \in [0,1]$. The spectral entropy of the signal is then computed as the Shannon entropy of this distribution, $E_S(s)$, where $\varepsilon$ is a small constant for numerical stability[8]:

$$E_S(s) = -\sum_{f} \widetilde{P}_s(f) \log_2\left(\widetilde{P}_s(f) + \varepsilon\right) \tag{4}$$

The spectral entropy loss, $\mathcal{L}_{SE}$, is then defined as the mean squared error between the spectral entropies of the predicted and ground truth signals (Equation 5). This term encourages the model to preserve the spectral characteristics of the true ECG tissue.

$$\mathcal{L}_{SE} = \frac{1}{B}\sum_{i=1}^{B}\left(E_S(y_i) - E_S(\hat{y}_i)\right)^2 \tag{5}$$

To also ensure morphological accuracy in the time domain, we include a Huber loss term, $\mathcal{L}_H$, which combines the robustness of mean absolute error with the smoothness of mean squared error. Hence, total loss, $\mathcal{L}_{total}$, is a weighted sum of $\mathcal{L}_H$ and $\mathcal{L}_{SE}$:

$$\mathcal{L}_{total} = \mathcal{L}_H + \omega(n) \cdot \mathcal{L}_{SE} \tag{6}$$

The weight function for spectral entropy loss term, $\omega(n)$, is defined by a cosine decay schedule, where $n$ is the current epoch and $E$ is the total number of training epochs.

$$\omega(n) = \frac{1}{2}\left(1 + \cos\left(\frac{\pi n}{E}\right)\right) \tag{7}$$

This schedule emphasises spectral differences in the early stages of training and gradually shifts focus to time-domain accuracy, allowing for initial spectral regularisation without constraining final optimisation. The dataset was split into training (80%, 240 subjects), validation (10%, 30 subjects), and test (10%, 30 subjects) sets, with class distributions balanced across all splits. The model was trained for 200 epochs using the Adam optimiSer with an initial learning rate of 0.001, which was halved if the validation loss plateaued for five consecutive epochs. The model state with the highest validation $R^2$ score was selected as the best model.

## 3. Results

Our model closely matched the ground truth (Figure. 2), achieving a mean $R^2$ of $0.99 \pm 0.01$ on the hold-out test set.

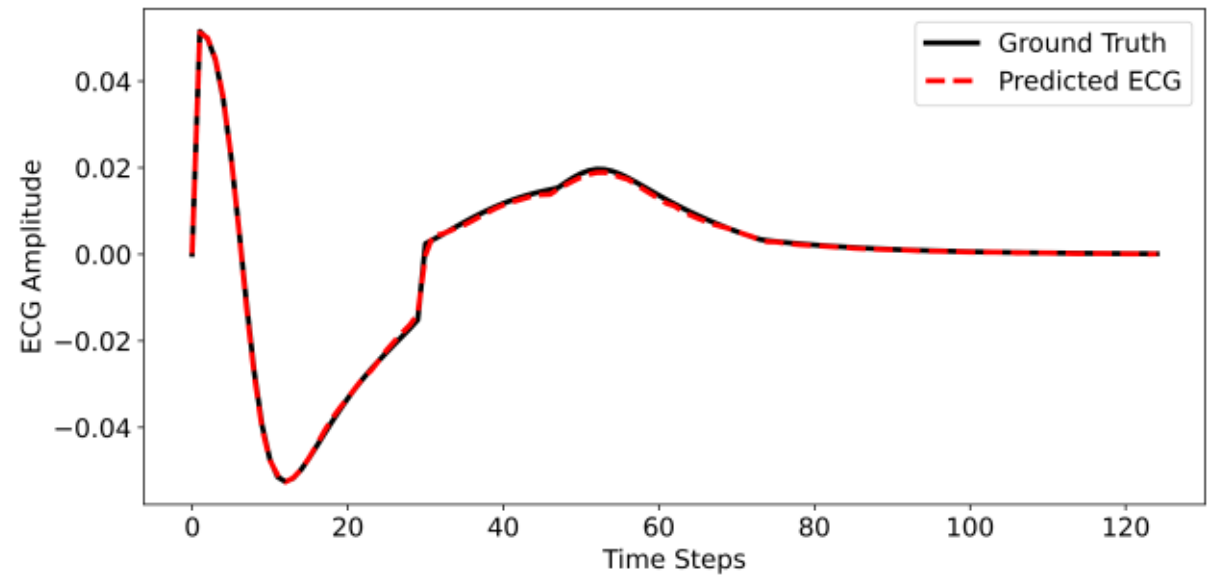

**Figure 2.** Test set example of predicted ECG (red dashed line) and ground truth (black line).

Performing a model ablation study (Figure 2) revealed that removing any model component reduced performance. While excluding the spectral entropy loss did not significantly impact results for fibrotic or gap junction-remodelled tissue ($p > 0.05$), it improved $R^2$ and MAE in healthy tissue ($p < 0.05$). This indicates that the spectral entropy loss specifically enhances accuracy in homogeneous tissue, even when the training set is dominated by heterogeneous cases. Note in experiments with no convolutional neural network (CNN), the mean voltage was evaluated as the surrogate for encoder latent space vector.

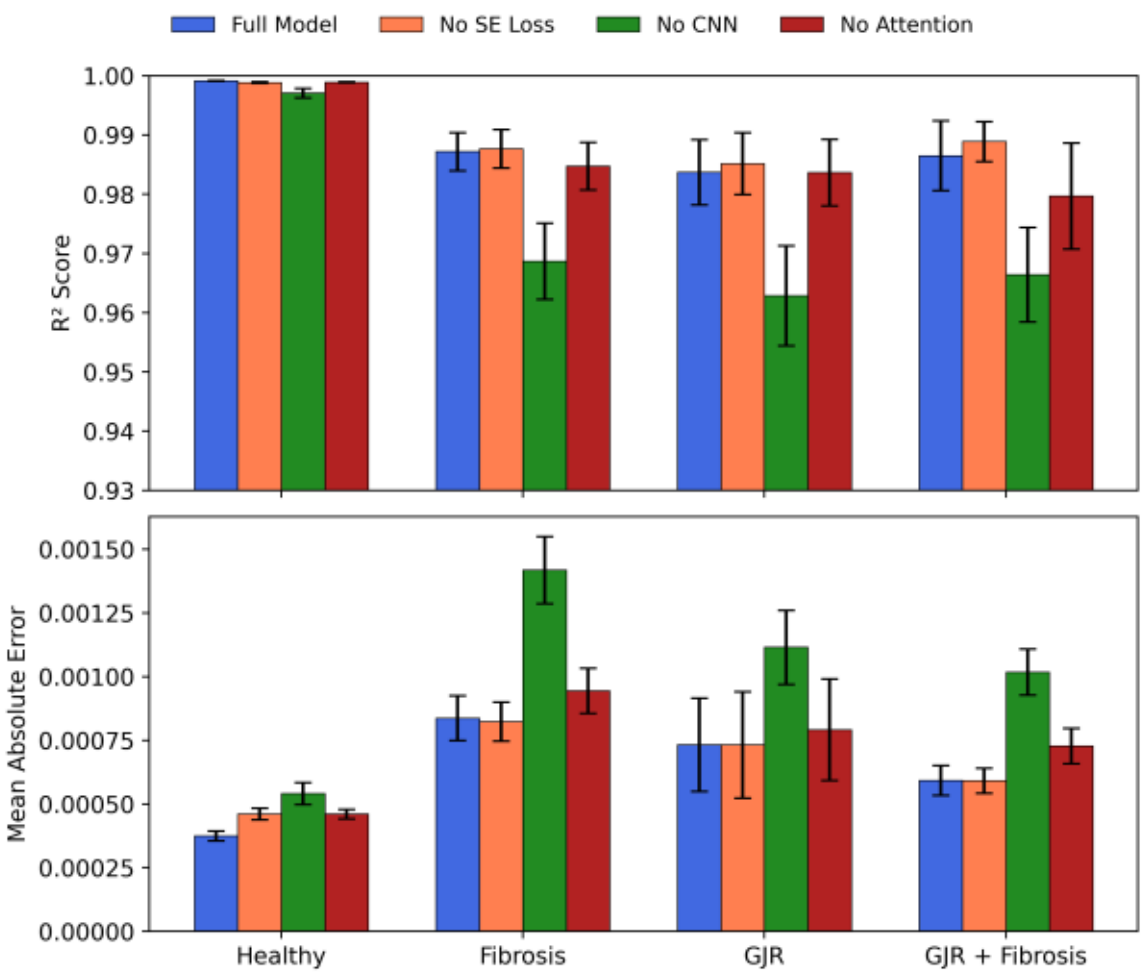

**Figure 3.** Bar charts of $R^2$ score and mean absolute error of the ablation experiments of model components and loss function across all tissue conditions in the test set.

## 4. Discussion

This study investigates the application of DL to solve the forward problem in electrocardiology. Our results demonstrate strong agreement between DL predictions and in-silico ground truth on two-dimensional voltage propagation data. The proposed DL model employs an

encoder–decoder architecture that leverages attention mechanisms and time embeddings to capture temporal dynamics, alongside a CNN to learn spatial relationships. We further introduce a novel application of spectral entropy loss function, which enhances the model's predictive performance under both homogeneous and inhomogeneous tissue conditions. To the best of our knowledge, this is the first study to apply DL to this specific problem. Overall, our findings highlight that DL can effectively map voltage propagation patterns to extracellular signals. The key advantage of this approach lies in the reduced computational cost and the parallel processing capabilities of DL, which could improve the efficiency of technologies relying on forward solutions, such as cardiac digital twin calibration and electrocardiographic imaging. A potential future direction is to train the model directly on experimental data. Traditionally, it has been assumed that achieving an accurate forward solution requires incorporating inhomogeneities into the calculations, as in the bidomain model. However, Bear et al. demonstrated that including torso inhomogeneities reduced the forward solution error compared to a homogeneous torso solution (the ground truth in our study) but did not fully resolve discrepancies when compared to measured body surface potentials [9]. A potential future direction of this research is to investigate whether a DL-based approach, when trained solely on experimental data, could surpass the accuracy of the current state-of-the-art forward solution based on the bidomain equation. Which in turn can improve accuracy of inverse solution, as the forward problem is integral to the current solution [10–12]. Future research should also explore extending this work to three-dimensional, patient-specific cardiac geometries, as well as predicting ECGs from bidomain simulations or experimental data to assess model performance in capturing signals from inhomogeneous torsos.

## 5. Conclusion

This study shows that DL can accurately solve the forward problem in electrocardiology, matching in-silico ground truth from 2D voltage propagation. Our encoder–decoder model, enhanced with attention, time embeddings, and a spectral entropy loss, performed well in both homogeneous and inhomogeneous tissue. These results position DL as a promising alternative to physics-based solvers.

## Acknowledgments

This work was supported by the British Heart Foundation (PG/23/11552).

**Address for correspondence:**
Dr. Shaheim Ogbomo-Harmitt.
Floor 3 Lambeth Wing, St Thomas' Hospital, London, SE1 7EH, UK.
shaheim.ogbomo-harmitt@kcl.ac.uk.